\documentclass[reprint,amsmath,amssymb,prl,aps,twocolumn]{revtex4-2}

\usepackage{graphicx}

\usepackage{bm}
\usepackage{amsmath}
\usepackage{amssymb}
\usepackage[latin1]{inputenc}
\usepackage{color}
\usepackage[colorlinks=true, pdfstartview=FitV, linkcolor=blue, citecolor=blue, urlcolor=blue]{hyperref} 
\usepackage{dcolumn}

\begin{document}

\title{Dislocation-driven relaxation processes at the conical to helical phase transition in FeGe}

\author{P. Schoenherr,$^{1,2,3}$, M. Stepanova,$^{4}$ E. N. Lysne,$^{4}$ N. Kanazawa,$^{5}$ Y. Tokura,$^{5,6}$ A. Bergman,$^{7}$ D. Meier$^{4,8}$} 

\affiliation{ $^{1}$Department of Materials, ETH Z\"urich, Vladimir-Prelog-Weg 4, 8093 Zurich, Switzerland
 $^{2}$School of Materials Science and Engineering, UNSW Sydney, Sydney, NSW 2052, Australia\\
 $^{3}$ARC Centre of Excellence in Future Low-Energy Electronics Technologies (FLEET), UNSW Sydney, Sydney, NSW 2052, Australia\\
 $^{4}$Department of Materials Science and Engineering, Norwegian University of Science and Technology (NTNU), Sem S\ae landsvei 12, 7034 Trondheim, Norway\\
 $^{5}$Department of Applied Physics, University of Tokyo, Tokyo 113-8656, Japan\\
 $^{6}$RIKEN Center for Emergent Matter Science (CEMS), Wako 351-0198, Japan\\
 $^{7}$Department of Physics and Astronomy, Uppsala University, PO Box 516, Uppsala 75120, Sweden\\
 $^{8}$Center for Quantum Spintronics, NTNU, Trondheim, Norway
}

\date{\today}

\begin{abstract}

The formation of topological spin textures at the nanoscale has a significant impact on the long-range order and dynamical response of magnetic materials. We study the relaxation mechanisms at the conical-to-helical phase transition in the chiral magnet FeGe.  By combining ac susceptibility, magnetic force microscopy measurements and micromagnetic simulations, we demonstrate how the motion of magnetic topological defects, here edge dislocations, impacts the local formation of a stable helimagnetic spin structure. Although the simulations show that the edge dislocations move with a velocity of about $100$\,m/s through the helimagnetic background, their dynamics are observed to disturb the magnetic order on the timescale of minutes due to pinning by randomly distributed structural defects. The results corroborate the substantial  impact of dislocation motions on the nanoscale spin structure in chiral magnets, revealing previously hidden effects on the formation of helimagnetic domains and domain walls.

\end{abstract}

\maketitle
\section{Introduction}

\begin{figure*}
	\centering
	\includegraphics[width=0.75\textwidth]{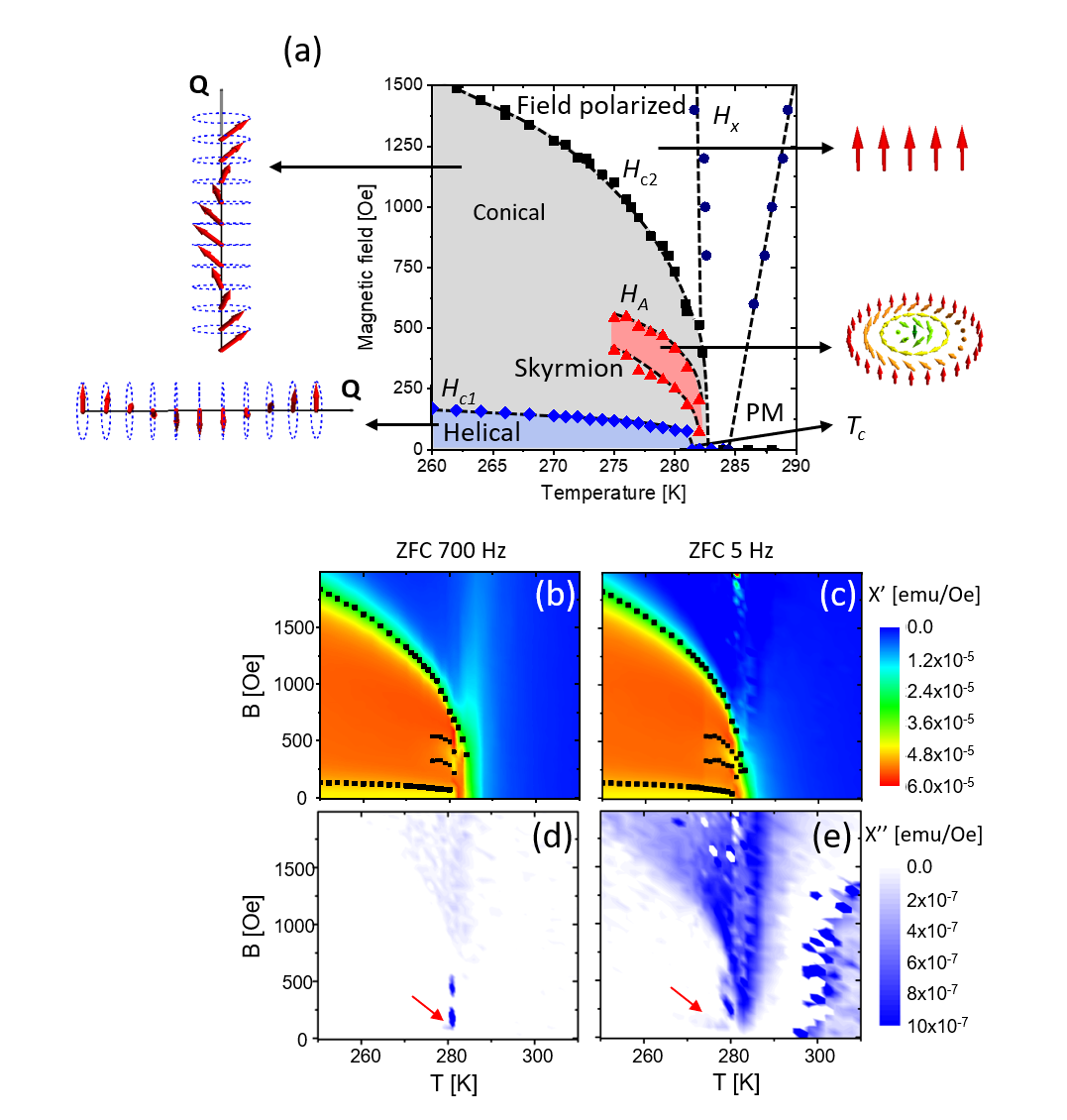}
	\caption{(a) Magnetic phase diagram of FeGe with $H \parallel$ [100] deduced from ac susceptibility measurements. The graph shows five magnetic phases (paramagnetic phase (PM), helimagnetic phase, conical phase, skyrmion phase and the field polarised state; different colours for data points are used for better visibility) with the corresponding magnetic structures sketched next to the phase diagram. For further detail on how the phase boundaries are determined see Supplementary Methods. (b)-(e) Contour plots for the real $\chi'$ ((b) and (c)) and imaginary $\chi''$((d) and (e)) magnetic susceptibility for frequencies of 700\,Hz and 5\,Hz (decreasing magnetic field). The black squares indicate the phase boundaries and the red arrows mark the location of the conical-to-helical phase transition. \label{edge1} }
\end{figure*}

Chiral magnetic materials exhibit a variety of topological structures, such as skyrmions, dislocations, disclinations and domain walls~\cite{Nagaosa2013, Nattermann2014, Schoenherr2018, Goebel2020}. Topological structures are promising candidates for new spintronics applications, where their stability and dynamical properties are utilized to design, e.g., next-generation memory technology~\cite{Fert2017} and artificial synapses for neuromorphic computing~\cite{Song2020}. Such non-trivial magnetic patterns occur in a wide range of materials including the group of non-centrosymmetric B20 materials~\cite{Muehlbauer2009, Nagaosa2013, Fert2013, Nattermann2014}. 
In these materials, the helimagnetic spin structures are stabilised by an interplay of ferromagnetic exchange and Dzyaloshinskii-Moriya (DM) interaction, representing the ground state as illustrated in Fig.~\ref{edge1}(a). Under the application of a small magnetic field, the helical axis, described by the wave vector $\bf Q$, rotates in the direction of the field and the spins form a conical spiral. For sufficiently high fields, all spins are fully aligned in the direction of the magnetic field (field-polarised state). The transition between the conical and helical phase is accompanied by complex relaxation mechanisms~\cite{Plumer1990, Bauer2017, Dussaux2016, Milde2020}, for which the formation and annihilation of topological defects play a crucial role~\cite{Milde2013}.

In order to structure the discussion, we distinguish three relaxation processes at the conical-to-helical phase transition. Each process is associated with a different time scale: (i) single spin dynamics $<$ 0.1\,ms, (ii) reorientation of the wave vector $\bf Q$ $\approx$ 1\,s~\cite{Bauer2017, Milde2020}, and (iii) slow relaxations via moving magnetic defects $\gg$ 1\,min~\cite{Dussaux2016}. It is important to note that the three processes are not completely independent. The movement of individual dislocations, for example, can cause variations in Q at the local scale, reflecting the complexity of the overall relaxation process.

Mechanism (i) and (ii) originate from spin-spin interactions and the transition between magnetic single- and multi-domain states, respectively. The long time scales of mechanism (ii) are proposed to result from the formation and destruction of domain walls~\cite{Bauer2017, Qian2016, Milde2020} that can possess a non-trivial topology~\cite{Schoenherr2018}. In contrast, relaxation mechanism (iii), with even longer time scales, was related to the  unpinning and subsequent motion of magnetic edge dislocations. The latter was concluded indirectly on the basis of the observation of jump-like 180$^{\circ}$ degree phase shifts in spatiotemporal MFM maps and NV magnetometry data gained on FeGe~\cite{Dussaux2016}, that is,  without resolving the moving defects. Thus, the actual defect motion and its impact on the local and global magnetic structure formation, e.g., domains and domain walls, remain to be demonstrated.

Here, we combine frequency-dependent ac susceptibility measurements, magnetic force microscopy (MFM) and micromagnetic simulations to investigate the temporal evolution of the chiral magnetic order in FeGe across the conical-to-helical phase transition. We focus on studying edge-dislocation-related magnetic relaxation processes in FeGe on both microscopic and macroscopic length scales. Spatiotemporal MFM maps provide direct evidence for the motion of magnetic edge dislocations and reveal how moving dislocations interfere with the formation of stable domains and domain walls in the helimagnetic state. The simulations indicate that individual dislocations in FeGe have a velocity of about $80$\,m/s. The unpinning of edge dislocations, however, is a statistical process, disturbing the local magnetic structure up to hours after the magnetic-field removal. Despite the severe impact on the spin system at the nanoscale, such micromagnetic effects are hidden on macroscopic length scales as reflected by our ac susceptibility data.\\

The phase diagram and magnetically ordered states of a [100]-oriented FeGe crystal (see Supplementary Methods) are presented in Fig.~\ref{edge1}(a). Phase boundaries are derived from ac susceptibility measurements (1 Oe, 700 Hz, H $\parallel [100]$). Below the magnetic ordering temperature $T_c = 281.4 \pm 0.2$\,K (determined from the phase diagram in Fig.~\ref{edge1}(a),  FeGe displays helimagnetic long-range order as sketched in Fig.~\ref{edge1}(a). Application of a magnetic field induces a conical spin structure at $H_{c1}$ and a field-polarized state for $H > H_{c2}$. $H_A$ indicates the transition to the well-established skyrmion phase in FeGe and $H_x$ marks the transition into the paramagnetic state determined by a fluctuation-disorder regime~\cite{Janoschek2013}. The measured phase diagram is in good agreement with literature~\cite{Yu2010, Bauer2012, Bauer2016, Wilhelm2011, Cevey2013, Turgut2017}, representing the basis for our magnetic relaxation investigations. 
\begin{figure*}[t]
	\centering
	\includegraphics[width=1\textwidth]{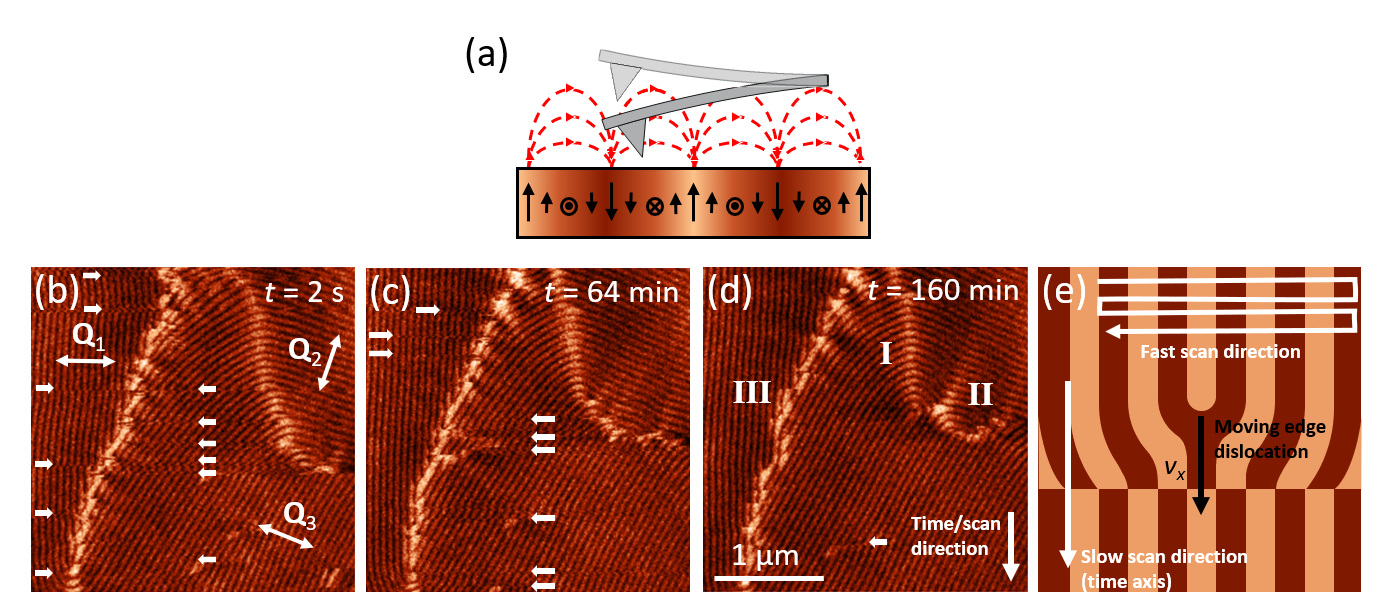}
	\caption{(a) Schematic sketch of the working principle of an MFM measurement on a
		FeGe sample.  (b)-(d) MFM image at 269\,K 2\,s, 64\,min and 160\,min after removing the magnetic field (the magnetic field was applied perpendicular to the surface normal). The images show an area of three {\bf Q} vector orientations separated by three domain walls, namely a curvature wall (I), a zig-zag disinclination wall (II) and an edge dislocation wall (III). The white arrows indicate $180^{\circ}$ phase jumps in the MFM images. (e) Skematic sketch of an MFM image from a helimagnetic structure with an edge dislocation unpinning and overtaking the magnetic tip during its scanning process. As the edge dislocation moves ($v_x$) by the magnetic pattern changes from one scan line to the next shifting the helimagnetic structure by 180$^{\circ}$.\label{edge2}}
\end{figure*}

In order to analyse transition-specific relaxation times, we consider frequency-dependent susceptibility measurements as introduced in Ref~\cite{Levatic2014} for Cu$_2$OSeO$_3$. Fig.\ref{edge1}(b)-(e) represent contour plots, showing the real and imaginary susceptibility --  $\chi'$ and $\chi''$, respectively -- for two frequencies, $f =$ 700\,Hz and $f =$ 5\,Hz , recorded with decreasing magnetic field (H $\parallel [111]$) (for $f =$ 0.1\,Hz see Supplementary Information). The phase transitions identified by the real part of the susceptibility in Figs.~\ref{edge1}(b)-(c) are in agreement with the phase diagram in Fig.~\ref{edge1}(a), indicating the same behaviour for magnetic fields along the [100] and [111] direction. Furthermore, the temperatures and magnetic fields at which phase transitions occur are largely frequency-independent aside from subtle changes around the skyrmion phase and the $H_x$ transition, as well as a small decrease in $T_c$ for smaller frequencies. The frequency-dependent decrease in $T_c$ indicates that slowly varying magnetic fields destabilize the helimagnetic order, which is similar to earlier findings on Cu$_2$OSeO$_3$~\cite{Levatic2014}. In contrast, the imaginary part $\chi''$ exhibits a strong frequency dependency as seen in Figs.~\ref{edge1}(d)-(e), indicating significant time-dependent  dissipation processes at the different magnetic phase transitions. At high frequencies (700\,Hz, Fig.~\ref{edge1}(d)), we record pronounced signals related to the skyrmion phase and the field-aligned state; these signals increase even further towards low frequencies (5\,Hz, Fig.~\ref{edge1}(e)). This signal increase towards lower frequencies indicates slow magnetic relaxations in the order of a few seconds~\cite{Levatic2014, Qian2016, Bannenberg2016, Bauer2012}.

Most importantly for this work, the imaginary susceptibility $\chi''$ shows a signal related to the conical-to-helical transition, which becomes more pronounced towards lower frequencies as indicated by the red arrows in Fig.~\ref{edge1}(d) and (e). The respective relaxation time can be extracted from frequency-dependent ac-susceptibility measurements by simultaneously analysing $\chi'$ and $\chi''$ by a modified Cole-Cole formalism~\cite{Qian2016, Bannenberg2016}. In the evaluated temperature regime (269\,K-280\,K), we find relaxation times below 0.2\,s (see Supplementary Information for details). The observation of a relatively long relaxation time is consistent with other $B20$ materials and is associated with the reorientation of the {\bf Q} vector~\cite{Bauer2017, Milde2020}. Reported time scales range from milliseconds~\cite{Qian2016} to tens of seconds~\cite{Bannenberg2016}.

Thus, our macroscopic ac-susceptibility measurements suggest that in the temperature range of 269\,K-280\,K FeGe relaxes into the helimagnetic ground state in about 1\,s. According to the established relaxation mechanisms, the relaxation includes the processes (i) and (ii). In contrast, the longer relaxation times associated with moving edge dislocations (iii)  is not resolved in the macroscopic measurements. Thus, we perform additional spatially resolved MFM measurements to investigate the local micromagnetic dynamics and corresponding time scales.\\

To study the micromagnetic relaxations that occur as FeGe enters its helimagnetic ground state across the conical-to-helical phase transition, we proceed as follows. First, the system is exposed to a magnetic field of 120\,mT to induce the field polarized state (in-plane magnetic field). Then, after removing the magnetic field, MFM scans are recorded to analyse the local relaxation dynamics (Fig.~\ref{edge2}(b)-(d)). As images are collected line by line, the data provides both spatial and temporal resolution, with the latter being limited by the time needed to record an individual scan line ($\approx 7.5$\,s for dual-path MFM  as applied here). Figures~\ref{edge2}(b)-(d) thus represent spatio-temporal maps with time, $t$, evolving along the slow scan direction (vertical direction). The images are taken 2\,s, 64\,min and 160\,min after removing the magnetic field. The helimagnetic phase shows up as bright and dark lines (perpendicular to ${\bf Q}$) in the MFM images originating from the stray field of the magnetic structure (Fig.~\ref{edge2}(a)).

Right after removing the magnetic field, three areas with uniform ${\bf Q}$-vector direction (${\bf Q}$ domains) are observed in Fig.~\ref{edge2}(b) denoted as ${\bf Q}_1$-${\bf Q}_3$ . The domains are separated by different domain walls, including all fundamental types, namely one curvature wall (type I, marked in Fig.~\ref{edge2}(d)), one zig-zag disclination wall (type II) and one edge dislocation wall (type III). For a detailed coverage of the different types of helimagnetic domain walls in FeGe, we refer to ref~\cite{Schoenherr2018}. The image corroborates that the 70~nm periodicity and direction of $\bf Q$ (relaxation process (ii)), as well as associated domain walls, are established immediately after removing the magnetic field ($t <$ 2\,s, Fig.~\ref{edge2}(b)). The inner structure of the domain walls is evolving over much longer timescales. This is linked to the $180^{\circ}$ phase jumps, which are observed even hours after removing the magnetic field (indicated by white arrows in Fig.~\ref{edge2}). Such $180^{\circ}$ phase jumps are the fingerprint of unpinning and subsequent climbing edge-dislocations with the number of jumps decaying with $1/t$ as the system relaxes~\cite{Dussaux2016} (see Fig.~\ref{edge2}(e) for an illustration). Figure~\ref{edge2}(b)-(d) shows that edge-dislocation triggered phase jumps extend over several $\mu$m, even in the presence of domain walls. Importantly, these moving edge dislocations also carry a topological charge of $1/2$~\cite{Schoenherr2018}, which alters the topological structure of the domain walls whenever a defect gets pinned by the wall. In case an edge dislocation gets pinned at a domain wall, it will add $1/2$ to the overall domain wall charge and alter its topology. This leads us to the conclusion that dislocation dynamics strongly impacts the formation of stable domains and domain walls, leading to ultra-slow build-up times for domain walls as seen in Fig.~\ref{edge2}(b)-(d). 

\begin{figure}[t]
	\centering
	\includegraphics[width=0.4\textwidth]{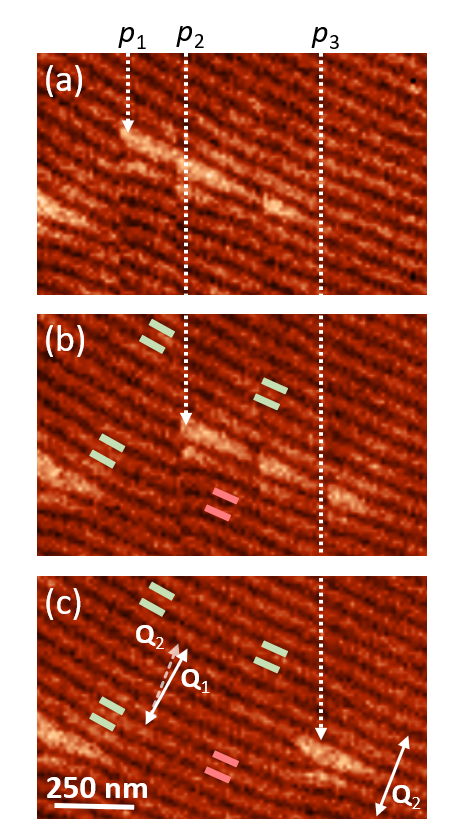}
	\caption{(a)-(c)  MFM image series (269\,K) showing edge dislocations at position $p_1$, $p_2$ and $p_3$. Along with the position of the edge dislocations denoted by $p_2$ to $p_3$, a $180^{\circ}$ shift is induced in the magnetic structure, which is indicated by the two pink lines. The green lines highlight regions that are not effected by the edge dislocation movement. The white arrows indicate the {\bf Q} vector direction before and after the edge dislocation. \label{edge3}}
\end{figure}

Due to the significant impact of the moving edge dislocations on the development of the local helimagnetic structure,  it is important to understand their dynamical properties. Direct imaging of the edge dislocation motion by MFM is not feasible because of the limited scan speed.  On rare occasions, however,  snapshots of mobile dislocations can be obtained as shown in Fig.~\ref{edge3}. Figure~\ref{edge3} presents an MFM image series with edge dislocations pinned at three positions $p_1$ to $p_3$, corroborating that edge dislocations can climb through the helimagnetic structure. As edge dislocations change position, they induce a 180$^{\circ}$ shift in the magnetic structure (see pink lines in Figs.~\ref{edge3}(b),(c)), consistent with what is seen in Fig.~\ref{edge2}. This finding corroborates the previous assumption that a moving edge dislocation induces 180$^{\circ}$ phase shift in the helimagnetic structure. Further can be noticed that the direction of the helimagnetic structure changes by approximately $3^{\circ}$ when going across the edge dislocation (see white arrows indicating the {\bf Q} vector direction in Fig.~\ref{edge3}(c)). The fact that dislocations are resolved in three consecutive scans lets us assume that they are pinned by structural defects, prohibiting their free motion. Thus, the MFM data in Fig.~\ref{edge3} alone does not allow for drawing conclusions about the dislocation motion itself.\\

\begin{figure*}
	\centering
	\includegraphics[width=1\textwidth]{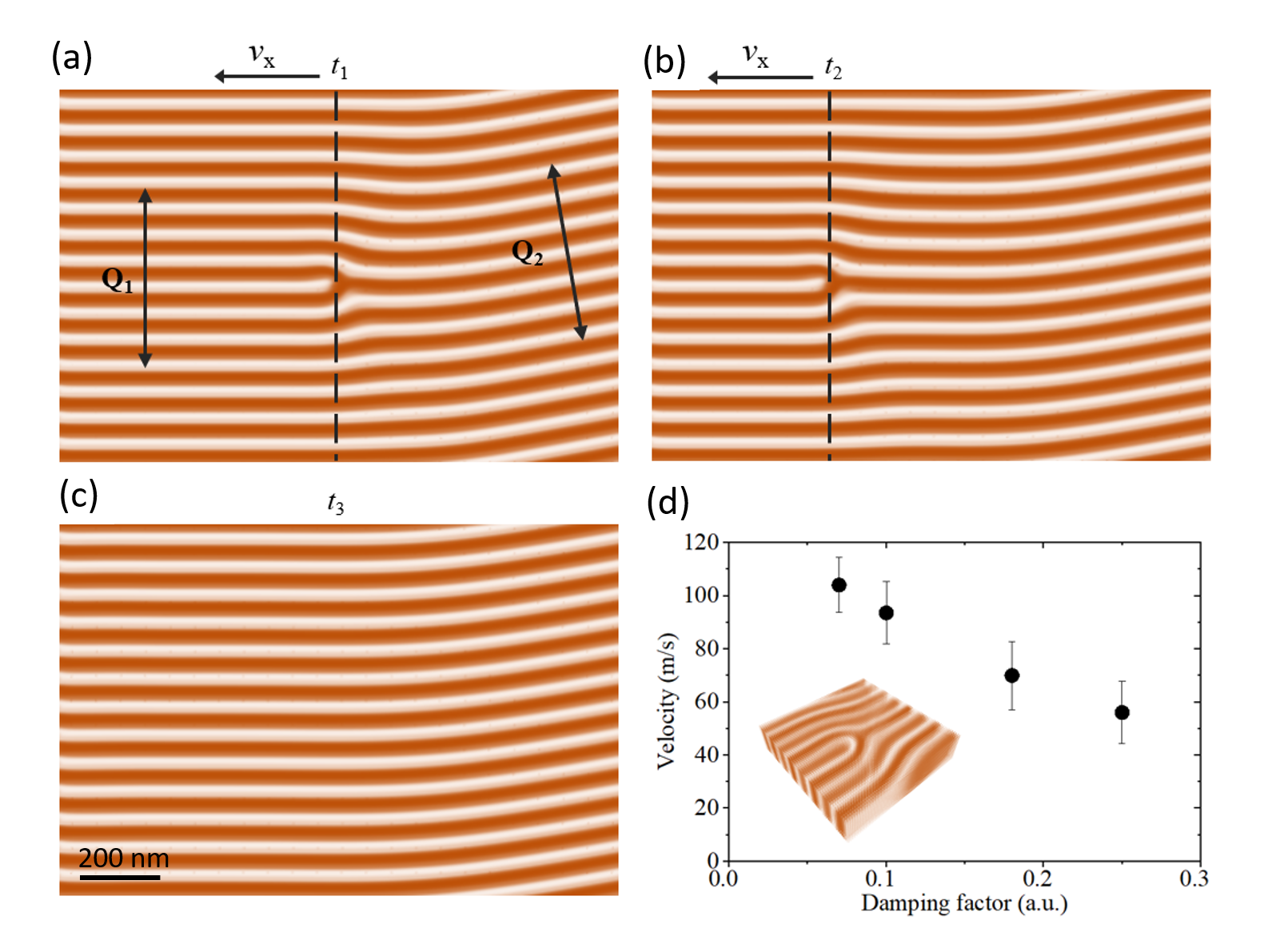}
	\caption{Micromagnetic simulations modelling the motion of a magnetic edge dislocation in the helimagnetic phase of FeGe. (a),(b) and (c) are snapshots from the time-dependent simulation, illustrating the motion of a dislocation that initially formed between the two regions described by the wave vectors ${\bf Q}_1$ and ${\bf Q}_2$ ($\angle({\bf Q}_1,{\bf Q}_2) = 10$\,degree). The black dashed line indicates the position of the dislocation and $v_x$ denotes the direction of motion (damping factor $\alpha = 0.25$, thickness $d = 32$\,nm; see main text for details). (d) Relation between the edge dislocation velocity and the damping factor  $\alpha$. Error bars account for the unknown thickness $d$ of the surface domain through which the edge dislocation is moving, varying between $d = 8$\,nm and $64$\,nm in our simulations. The average velocity for the investigated regime is $v_x = 80 \pm 20$ m/s. \label{edge4}}
\end{figure*}

To gain additional insight and understand the dynamics of the edge dislocations in FeGe, we perform 3D micromagnetic simulations and model their relaxation dynamics using MuMax$^3$~\cite{Vansteenkiste2014}. Figure~\ref{edge4}(a)-(c) displays three representative snapshots from the time-dependent simulations, showing how an edge dislocation climbs through the helimagnetic background. The simulations are performed for a volume of 2048 $\times$ 1024 $\times$ 32 nm$^3$ and a damping factor $\alpha = 0.25$~\cite{Beg2017}.  Consistent with the experimental data in Fig.~\ref{edge3}(c), it is sufficient to introduce a small variation in the orientation of ${\bf Q}$ to achieve the edge dislocation in the initial state (Fig.~\ref{edge4}(a)). In this example, the orientation differs by $10^{\circ}$ between the left and the right part of the simulated volume described by ${\bf Q}_1$ and ${\bf Q}_2$, respectively. We find that as the helimagnetic structure relaxes, the dislocation climbs about $260$\,nm in $4$\,ns (Fig.~\ref{edge4}(a),(b)), which corresponds to a velocity of $\approx 65$\,m/s. The gain in energy associated with the ejection of the dislocation can be estimated by comparing the data in Fig.~\ref{edge4}(a) and (c), indicating a reduction of about $400$\,J/m$^3$. A systematic analysis of the dislocation velocity as function of the damping factor is presented in Fig.~\ref{edge4}(d), considering the range of $\alpha$ values reported for FeGe in literature~\cite{Turgut2017,Beg2017}. In addition, to account for the unknown thickness of the magnetic surface domains in FeGe~\cite{Schoenherr2018} we simulate the dynamics for different realistic surface-domain thicknesses $d$ between 8\,nm and 64\,nm, which leads to the error bars in Fig. ~\ref{edge4}(d). The result shows that the dislocations slow down as the damping increases, yielding an average velocity of $v_x = 80 \pm 20$\,m/s for the investigated regime, which is comparable to the velocities reported for ferromagnetic domain walls~\cite{Miron2011}. We note that the calculated direction of the dislocation motion is opposite to ref.~\cite{Dussaux2016}, which we ascribe to methodological differences in how the DM interactions and boundary conditions were treated. However, as the 3D surface domain structure is unknown and we do not expect an impact on the calculated velocity, this aspect is not investigated further. Thus, the micromagnetic simulations demonstrate that edge dislocations in a defect-free environment are moving on much faster timescales than captured by our MFM experiments. This result corroborates the assumption that the dislocations in Fig.~\ref{edge3} are pinned by structural defects, which allows resolving them in different locations. \\

In conclusion, our ac susceptibility, MFM data and micromagnetic simulations provide insight into the relaxation processes happening at the conical-to-helical phase transition in FeGe. Our local results reveal the dynamics of magnetic edge dislocations in chiral magnets and the role these topological defects are playing for the relaxation of the nanoscale spin structure. We show that the timescales associated with the edge dislocation motion are rather fast with dislocation velocities in the order of $\approx 100$\,m/s. The magnetic edge dislocations, however, can get trapped by randomly distributed (structural) defects. Thus, the timescale is dominated by "unpinning" events, which is a statistical process, and explains the long relaxation times associated with the motion of the dislocations. Although the dislocation-driven relaxation process does not manifest in macroscopic measurements, it has a significant impact on the magnetic structure formation inhibiting the formation of locally stable spin arrangements, domains and  domain walls on the timescale of seconds to minutes. Thus, our results reveal the fundamental importance of moving edge dislocations for the relaxation dynamics of chiral magnets and lamella-like magnetic textures in general, demonstrating their substantial impact on the magnetic order at the nanoscale.

\textbf{Acknowledgement}
We thank L. K{\"o}hler and M. Garst for fruitful discussions about the edge dislocation and domain wall interaction and M. Trassin for introducing P.S. to squid measurements. We thank M. Fiebig for direct financial support. The work at ETH was supported by the Swiss National Science Foundation through grants 200021-149192, 200021-137520. N.K. acknowledges funding from JSPS KAKENHI (Grant number JP20H05155). Y.T acknowledges support from JST CREST (JPMJCR1874). D.M., M.S., and E.N.L. acknowledge support from the Research Council of Norway (FRINATEK project no.263228) and through its Centres of Excellence funding scheme, Project No. 262633, QuSpin. D.M. further thanks NTNU for support via the Onsager Fellowship Program and the Outstanding Academic Fellows Program.


\end{document}


\maketitle

\begin{affiliations} 
	\item Department of Materials, ETH Z\"urich, Vladimir-Prelog-Weg 4, 8093 Zurich, Switzerland
	\item School of Materials Science and Engineering, UNSW Sydney, Sydney, NSW 2052, Australia
	\item ARC Centre of Excellence in Future Low-Energy Electronics Technologies (FLEET), UNSW Sydney, Sydney, NSW 2052, Australia
	\item Department of Materials Science and Engineering, Norwegian University of Science and Technology (NTNU), Sem S\ae landsvei 12, 7034 Trondheim, Norway
	\item Department of Applied Physics, University of Tokyo, Tokyo 113-8656, Japan
	\item RIKEN Center for Emergent Matter Science (CEMS), Wako 351-0198, Japan
	\item Department of Physics and Astronomy, Uppsala University, PO Box 516, Uppsala 75120, Sweden
	\item Center for Quantum Spintronics, NTNU, Trondheim, Norway
\end{affiliations}

\newpage

\section{Methods}
	
	\textbf{Sample preparation and experimental details:} For our experiments we used FeGe (B20) single crystals grown by chemical vapour transport \cite{Richardson1967}. FeGe belongs to the P2$_1$3 materials and forms a helical spin spiral below T$_c = 278$\,K with a periodicity of 70\,nm \cite{Lebech1989}. To detect the local magnetic structure by MFM we used a standard NT-MDT NTEGRA Prima AFM in combination with a home-build cooling stage \cite{Dussaux2016}. MFM is a surface sensitive technique, which uses a magnetic cantilever detecting the out-of-plane stray field induced by the magnetic order. All images were taken with the same tip magnetisation in a temperature range of $265 - 270$\,K. Measurements were performed on (100) and (111)-oriented crystals that were aligned by Laue diffraction and cut in the desired direction. Afterwards the samples were chemo-polished to achieve a roughness below $\approx 1$\,nm in the MFM measurements. \\
	To compare the local magnetic behaviour with the macroscopic response, additional frequency dependent ac susceptibility measurements were conducted. The real $\chi'$ and imaginary $\chi''$ components of the susceptibility were measured on several FeGe crystals ($1 - 4$\,mg) in a Quantum Design MPMS3 SQUID. The DC magnetic field  was aligned with the [100] or [111] direction parallel to the AC field with an amplitude of 1 Oe. Field sweeps for increasing and decreasing magnetic field were measured at different frequencies (0.1\,Hz, 5\,Hz, 700\,Hz) after zero-field-cooling (ZFC). The phase boundaries ($H_{c1}$, $H_{c2}$, $H_A$) are derived from the extrema of $d\chi'/dB$ which corresponds to the infliction point of $\chi'$. $H_x$ is derived from the same measurement however by the infliction point of $\chi'$ vs $T$ \cite{Wilhelm2011}.
	
	\textbf{Micromagnetic simulations:} Micromagnetic simulations were performed using the open-source micromagnetic simulation framework MuMax$^3$ \cite{Vansteenkiste2014} (version: 3.10), based on the Landau-Lifshitz equation where contributions from the demagnetizing field were neglected. The simulations were performed at $T=0$ K using the following parameters for FeGe\cite{Beg2015}:  the saturation magnetization $M_s=384$ kAm$^{-1}$, the exchange stiffness $A=8.78$ pJm$^{-1}$ and the Dzyaloshinskii-Moriya interaction $D=1.58$ mJm$^{-2}$, which corresponds to a helix period of 70 nm. The simulation volume is 2048 nm $\times$ 1024 nm $\times$ $d$, where $d$ varies between 8 and 64 nm  as discussed in the main text. The unit cell volume is  4 $\times$ 4 $\times$ 4 nm$^3$. The dislocation was generated by simulating of two magnetic domains with the corresponding angles of the ${\bf Q}$ vector $\phi_1=90^{\circ}$ and $\phi_2=100^{\circ}$. Periodic boundary conditions were set in $y$-axis.

\section{Contour plots for $f = 0.1$\,Hz}

\begin{figure*}
	\centering
	\includegraphics[width=0.7\textwidth]{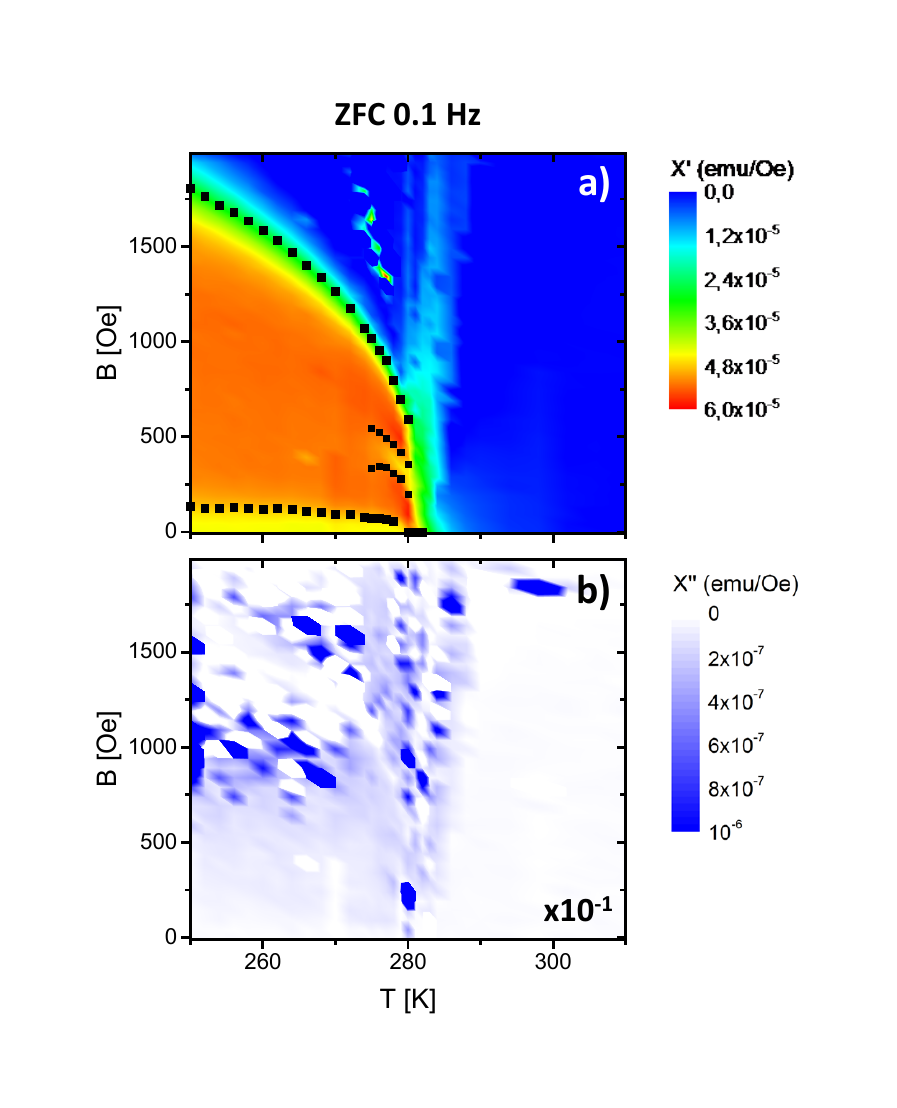}
	\caption{{\bf a} Contour plots for $\chi'$ {\bf b}  and $\chi''$ for a frequency of $0.1$\,Hz (decreasing magnetic field). The black squares indicate the phase boundaries.  \label{edge1} }
\end{figure*}

Figure~\ref{edge1} is showing the contour plots for a frequency of $f=0.1$\,Hz. The phase boundaries deduced from the real susceptibility are very similar to the $f=5$\,Hz and $f=700$\,Hz measurements. The decrease in transition temperature $T_c$ for slower frequencies which has been seen for the other two frequencies, can be seen for $f=0.1$\,Hz where it reaches $279$\,K. In contrast, the imaginary susceptibility is vastly different with an order of higher signal strength than in the faster experiments. However, no clear signals around the phase transitions and phases can be seen in comparison to the other measurements. Thus, we did not add it to the main text, but still want to show it for full disclosure in the supplementary. The increase in signal is clearly coming from the slower measuring speed, but this also seems to lead to higher noise and unreliable signals. 


\newpage

\section{Determination of relaxation times} 
At the conical to helical transition ($H_{c1}$) close to the transition temperature (270 - 280\,K) an increasing signal in $\chi''$ is seen in the 5 Hz phase diagram. To determine the relaxation times, we simultaneously analyse $\chi'$ and $\chi''$ by the modified Cole-Cole formalism~\cite{Bannenberg2016, Qian2016} ($\chi'$ and $\chi''$ are connected by the Kramers-Kronig relation, therefore reasonable values can only be found by analysing them together~\cite{Butykai2017}). 

\begin{figure*}
	\centering
	\includegraphics[width=0.9\textwidth]{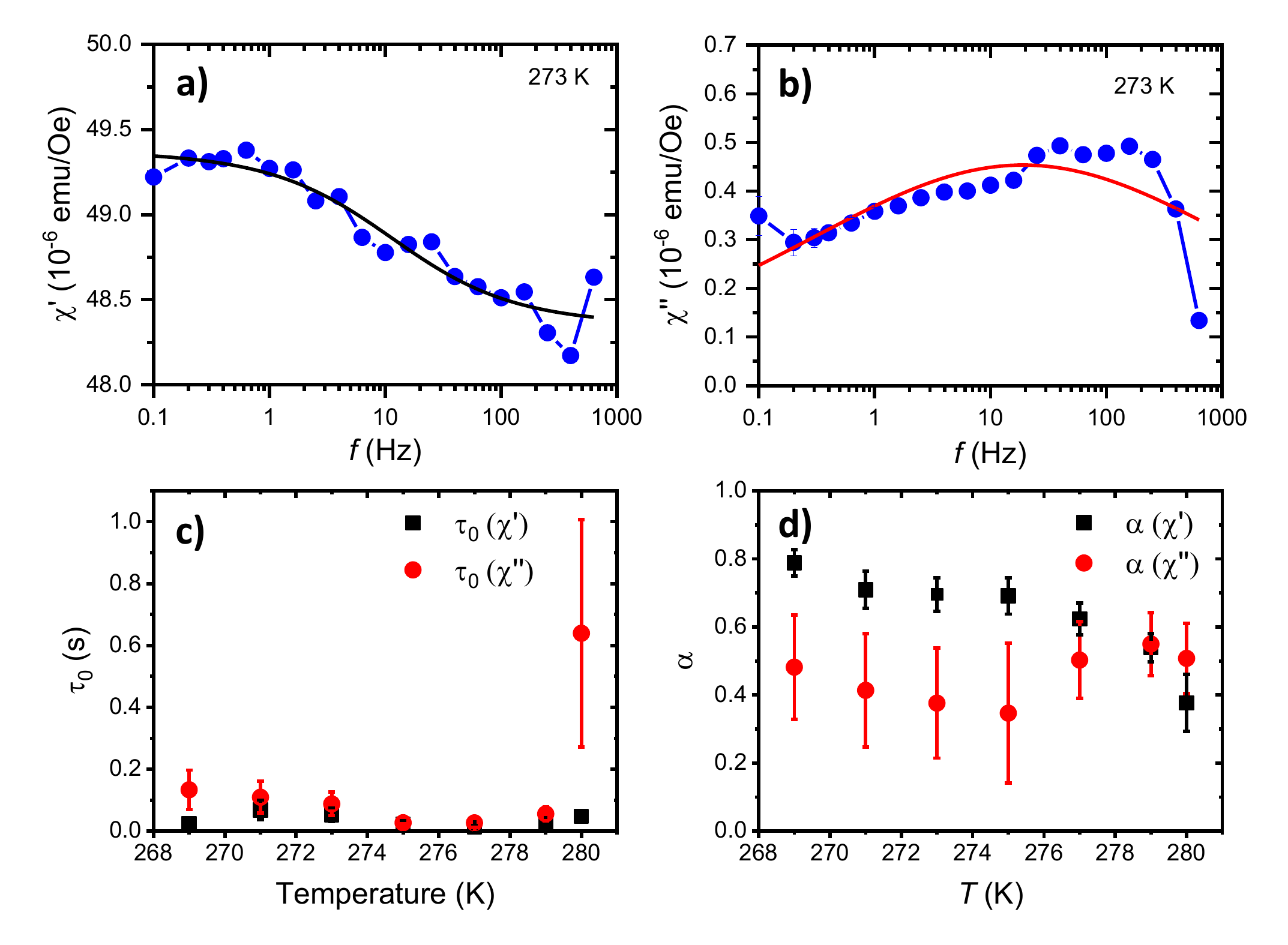}
	\caption{Fitting of the real {\bf a} and imaginary {\bf b} frequency-dependent susceptibility at 273\,K and under 10\,mT. {\bf c} Relaxation times and {\bf d} $\alpha$ determined by fitting the frequency dependent Squid measurements along the helical to conical phase transition (280\,K/4\,mT; 279\,K/6\,mT; 277\,K/8\,mT; 275\,K/9\,mT; 273\,K/10\,mT; 271\,K/11\,mT; 269\,K/11.5\,mT).\label{edge2} }
\end{figure*}

Modified Cole-Cole formalism:
\begin{equation}
\chi(\omega) = \chi(\infty) + \frac{\chi(0)-\chi(\infty)}{1+(i\omega\tau_0)^{1-\alpha}}
\end{equation}
$\chi(0)$ and $\chi(\infty)$ are the isothermal and adiabatic susceptibilities; $\omega$ is the angular frequency; and $\tau_0$ is the characteristic relaxation time. $\alpha$ is a parameter that defines the width of the relaxation frequencies distribution.  $\alpha=0$ corresponds to one single relaxation process and $\alpha = 1$ accounts for an infinitely broad relaxation distribution.
The equation can be separated into a real and imaginary part:
\begin{equation}
\chi'(\omega) = \chi(\infty) + \frac{A_0[1+(i\omega\tau_0)^{1-\alpha} \sin(\pi\alpha/2)]}{1+2(\omega\tau_0)^{1-\alpha}\sin(\pi\alpha/2)+(\omega\tau_0)^{2(1-\alpha)}}
\end{equation}
\begin{equation}
\chi''(\omega) = \frac{A_0(i\omega\tau_0)^{1-\alpha} \cos(\pi\alpha/2)}{1+2(\omega\tau_0)^{1-\alpha}\sin(\pi\alpha/2)+(\omega\tau_0)^{2(1-\alpha)}}
\end{equation}

with $A_0 = \chi(0)-\chi(\infty)$. The equations can be fitted to our frequency data that was acquired along the helical to conical phase transition. An example can be seen in Fig.~\ref{edge2}a. All relaxation time values as well as $\alpha$ can be seen in Fig.~\ref{edge2}b-c. Most of the relaxation times vary between 0.01\,s to 0.2\,s and $\alpha$ ranges from 0.3 to 0.8. The real and imaginary susceptibility should give the same relaxation times, which is true for almost all temperatures excluding 280\,K. This indicates that our Squid measurements are reliable and determine the macroscopic relaxation times. All $\alpha$ values are between 0-1 indicating that several relaxation processes are happening at this phase transition.

\newpage